\documentclass[manuscript]{acmart}
\usepackage{mathtools,halloweenmath}
\AtBeginDocument{%
  \providecommand\BibTeX{{%
    \normalfont B\kern-0.5em{\scshape i\kern-0.25em b}\kern-0.8em\TeX}}}

\acmConference[CHI '23]{CHI '23:  Workshop on
Generative AI and HCI}{April, 2023}{Hamburg, Germany}
\acmBooktitle{CHI '23: Workshop on
Generative AI and HCI,
  April, 2023, Hamburg, Germany}

\begin{document}

\title{Hidden Layer Interaction: A Co-Creative Design Fiction for Generative Models}

\author{Imke Grabe}
\email{imgr@itu.dk}
\author{Jichen Zhu}
\email{jicz@itu.dk}
\affiliation{%
  \institution{IT University of Copenhagen}
  \streetaddress{Rued Langgaards Vej 7}
  \city{Copenhagen}
  \country{Denmark}
  \postcode{2300}
}




\begin{abstract}

This paper presents a speculation on a fictive co-creation scenario that extends classical interaction patterns with generative models. 
While existing interfaces are restricted to the input and output layers, we suggest \textit{hidden layer interaction} to extend the horizonal relation at play when co-creating with a generative model's design space.
We speculate on applying feature visualization to manipulate neurons corresponding to features ranging from edges over textures to objects.
By integrating visual representations of a neural network’s hidden layers into co-creation, we aim to provide humans with a new means of interaction, contributing to a phenomenological account of the model’s inner workings during generation.

\end{abstract}
\maketitle

\section{Introduction}
A certain mystique surrounds the latent space of generative AI models. Recent art projects like the immersive installations from Refik Anadol's Machine Hallucination series\footnote{https://refikanadol.com/works/machine-hallucinations-nature-dreams/} illustrate how exploring latent images can be a captivating human experience. When humans are exposed to ``a world of the otherwise unseen''~\cite[p.1]{seberger_mystics_2020},
new phenomenological properties emerge, as~\citeauthor{seberger_mystics_2020} point out for the case of deepfakes. Aiming to make sense of this relation, \citeauthor{benjamin_machine_2021} apply post-phenomenology to analyze how machine learning models influence our ``phenomenological horizon''~\cite[p.12]{benjamin_machine_2021}. They introduce the concept of \textit{horizonal relations} to describe how a probabilistic machine learning model mediates between a human and the world. Knowing that feature visualization can reveal how parts of a neural network represent certain features, one might wonder how making sense of latent space's hidden properties could influence the horizonal relation coined by generative models when humans co-create with them.
%
While researchers have found ways to make sense of latent space's underlying properties in generative models~\cite{harkonen_ganspace_2020,bau_gan_2019}, 
its hidden layers are not a part of the interaction in co-creation~\cite{grabe_towards_2022}.
For example, one can find directions in latent space corresponding to semantic features~\cite{denton_image_2019,shen_interpreting_2020}, based on which humans might co-create with a generative model via features 
that translate back into the input vector~\cite{zaltron_cg-gan_2020}. Or, one can investigate how objects are encoded in GANs' internal representations~\cite{bau_gan_2019} serving as a backbone for rewriting their weights in an application that lets users rearrange objects at the output layer level~\cite{bau_rewriting_2020}. 
In both cases, however, human users interact with the input or output layer, respectively, which does not provide them with a sense of how changes feed back into the hidden layers.
Insight into a co-creator's actions matters as it informs our understanding of the decisions taken in the creative process. 
This opens up our speculation of whether insight into a generator's hidden workings could contribute to co-creative processes.

Design fiction can be useful for 
envisioning future interactions with generative AI~\cite{yildirim_emergent_2022,muller_drinking_2022}. 
In this paper, we use speculative design~\cite{auger_speculative_2013} as a way to imagine a co-creative scenario, that involves interaction with the hidden representations of a generative AI. 
More specifically, we imagine a reconstrained design~\cite{auger_reconstrained_2017} by combining existing technological elements, namely generative AI and feature visualization, to give human users a better sense of how a generative model's design space is constructed.
By asking 
\textit{How can interacting with hidden layers affect our experience of generative models?}, 
we aim to investigate whether 
feature visualization can expand the horizonal relation at play when co-creating with generative AI.


\section{Method}
Speculative design can serve as a method to overcome preconceptions and dogmas underlying technological development~\cite{auger_reconstrained_2017}. A characteristic of the recent development of generative AI is that we tend to compare the models' functionality to humans and expect similar behavior.
However, as AI models come with inherently different capabilities from humans, we risk restricting their development to one stringent direction without asking what novel modes of co-creation the technology might bring to the table. 
By recognizing the constraints underlying our thinking when imagining new technological applications, we might imagine `alternative presents' in a reconstrained world \cite{auger_reconstrained_2017}. 
One practical aspect differentiating generative models from human brains
is that one can `look inside' them by printing out the activation on its layers. However, these values are challenging to make sense of. Here, feature visualization can be used as a tool to shed light on what hidden layers react to.
By newly arranging technological elements \cite{auger_reconstrained_2017}, we speculate on a co-creation scenario that uses this insight.
In doing so, we apply an alternative motivation, namely to imagine a co-creator that gives us access to and lets us intervene with how its creation process is constructed.

\section{Current world: Closed Horizonal Relation}
Humans and generative AI can co-create following different patterns, 
ranging from simply prompting random generation over exploring design alternatives to manipulating the design space \cite{muller_mixed_2020,grabe_towards_2022,bau_rewriting_2020}. 
In the visual domain, latent variable models, such as Generative Adversarial Networks (GANs), use the input or output layer as the main access point for interaction. 
At the input level, humans can prompt GANs with a latent code that holds an encoding, such as a text description or other attributes~\cite{yildirim_disentangling_2018,zhu_be_2017}. Through interaction, alteration to the latent code lets humans traverse the model's design space~\cite{schrum_interactive_2020}. The generation process can also be influenced by intervening at the output layer level. \citeauthor{bau_rewriting_2020}~\cite{bau_rewriting_2020}~suggest rearranging elements in generated images to change the weights of the associated memory on the hidden layers. In other words, the output layer is the interface for making changes to the inside of the black-boxed network by rewriting its construction. 
%
In the terminology of evolutionary biology, one might say that we can either interact with a model via its phenotype, like in \citeauthor{bau_rewriting_2020}'s example, 
or via the genotype when manipulating the input vector, e.g., through encoded conditioning or by changing its variables. 
The layers in between remain undisclosed to the human co-creator. As generative models like GANs are probabilistic models and not rule-based, assuming their inner working is an impossible task.

\citeauthor{benjamin_machine_2021}~\cite{benjamin_machine_2021}~use post-phenomenology to analyze how humans might experience the interaction with probabilistic models, more specifically
how machine learning uncertainty functions as a design material. The authors formalize the phenomenological relationship by introducing \textit{horizonal relations}, depicted by the term ``ML $\sim$ World'', where the tilde operator describes the inference of the ML model from the world.
Interacting with machine learning models can then be formalized as ``Human $\Tilde{\rightarrow}$ Tech -- World'', 
where the arrow is the interpretation of the world based on the inferred model, through another technology such as the computer.
This shows how machine learning models can ```populate' the world that humans experience with ready-made yet uncertain entities''~\cite[p.12]{benjamin_machine_2021}. In other words, we read the world through the model's inference space~\cite{benjamin_machine_2021}. 
Applied to generative design, humans navigate in an inferred design space when co-creating with latent variable models.
With our fictive design application, we speculate on enriching the horizonal relation underlying the design space by
giving humans a sense of the hidden layers' workings by interacting through them.


\section{Speculative world: Hidden layer interaction}
In the following, we present our speculation on the fictive scenario of \textit{hidden layer interaction}, in which humans can edit the parameters inside a generative model via a visual interface. More specifically, we rearrange existing technological elements from the realm of computer vision and generative AI anew to come up with an alternative mode of interacting with a generative model.
Our motivation is to extend existing co-creation patterns by having humans interact with a model beyond its input and output layers. Central to our speculation, we suggest using feature visualization to investigate which parts of a neural network capture certain output features. Researchers have used the method to show how different layers in neural networks for image processing encode edges, textures, patterns, parts, and objects \cite{olah_feature_2017}.
Through optimization, we imagine visualizing what a particular neuron `sees.' In other words,
we could find neurons that activate strongly in connection to specific visual features.
These features then act as the visual representation of the linked neurons. By manipulating the visual representation linked to the neuronal activation, the user is imagined to `draw with neurons' in this activation space.

The interaction with the generative model is imagined as follows.
The user is presented with a selection of hidden layers of different abstraction levels ranging from edges to objects in the network.
They can choose to `pull up' one of those layers to change the neuronal activation on it.  
Here, they see facets that a feature captures, e.g., different geometric orientations for a low-abstraction layer representing edges, 
or different motives for a higher-abstraction layer representing objects.
When interacting with a chosen layer, the user can alter the strength of a feature facet via an interface similar to adjusting the exposure when editing an image. E.g., on a layer representing texture as a feature, they might increase the activation of some textures while decreasing the activation of others. The visual representation of the texture's facets acts as a handle to change the activation of the corresponding neurons.

Via these handles for visual features covering different levels of abstraction in the neural network, human users undertake internal modifications by changing the activation of neurons.
By observing how neuronal changes affect the output of a generative model, they experience the roles of neurons distributed across the network, what behavior they cause, and how they relate.
Hence, humans learn to think into the generative algorithm by looking backward in `time' into the generation process.
We argue that giving humans a sense of how a prompt at the input layer transforms through the multi-layered structure into an output affects how they experience the co-creative process by changing the horizonal relation towards the design space they navigate.
Integrating the inner workings of neural networks into co-creative processes lets the human user step inside their artificial co-creator.

%

\section{Conclusion}

We presented a co-creative generative design fiction making hidden representation experienceable in an interactive interface through feature visualization.
More specifically, we propose interacting through a generative models' inner workings in a future co-creation scenario, giving human actors a sense of how a generative AI's design space is constructed.
With our speculation, we aim to discuss the phenomenological relation underlying the experience of  co-creating with generative models.





\bibliographystyle{ACM-Reference-Format}
\bibliography{references.bib}


\end{document}